**Liminal Design: A Conceptual Framework and Three-Step Approach for Developing Technology that Delivers Transcendence and Deeper Experiences**

Johan Liedgren (1), Pieter Desmet (2) and Andrea Gaggioli (3,4)

(1) Independent researcher, Seattle, USA
(2) Department of Human-Centered Design, Delft University of Technology, Delft, The Netherlands
(3) Research Center in Communication Psychology, Università Cattolica del Sacro Cuore, Milan, Italy
(4) IRCCS, Istituto Auxologico Italiano - Milan, Italy

As ubiquitous technology is increasingly mediating our relationships with the world and others, we argue that the sublime is struggling to find room in product design primarily aimed at commercial and transactional goals such as speed and efficiency. We suggest a new category of products to promote deeper and more meaningful experiences, specifically those offering liminality, transcendence, and personal transformation. This paper introduces a conceptual framework and related three-step design approach that looks at narrative participation in design through abstractions to promote, hold and deepen more complex emotions. We explore implications from a theoretical point of view and suggest some liminal product design ideas as examples of how the model might be applied in practice.

**Introduction**

> *"It is one of the unexpected disasters of the modern age that our new unparalleled access to information has come at the price of our capacity to concentrate on anything much. The deep immersive thinking which produced many of civilization's most important achievements has come under unprecedented assault"* (De Botton, 2012, p. 264).

Social systems from the beginning of humanity have included activities and sacred places of contemplation in which to hold "liminal experiences": events that facilitate transcendence, interconnectedness, and feeling part of something larger than ourselves. The concept of liminality refers to the transitional phases in a human's life; phases that involve ambiguity and the dissolution of order that open a fluid or malleable space in which new ideas, practices and identities may emerge and develop (see e.g., Turner, 1974). The last twenty years' shift towards ubiquitous technology that mediates a lion's share of interactions with others and the world around us is instead focused on information sharing, ease of use, transactional speed, and platform integration. Users get two-dimensional renderings on a screen and words in messages, but rarely any deeper experiences beyond that.

In this paper, we argue that liminality is the missing design element for technology to better set the stage and open the door to real connections, more focus, and memorable experiences - all requiring some level of transcendence. For this, we need a new product-development framework looking beyond traditional user experience (UX) design and its predominant focus on instrumental, commercial, or pleasurable experiences. Liminal Design, by contrast, includes the perspective that any memorable, engaged, and meaningful participation with technology requires us to rethink interactions as social situations that also provide a safe path for the user's suspension of disbelief and the creation of a desire to consider and play with alternative narratives of the world - two prerequisites, we pose to experiences of transcendence. Such experiences will challenge targeted beliefs about ourselves and our reality, and thereby enable a deeply felt sense of personal disruption, surprise, and personal transformation.

In this context, we think about *liminality* as the space between two opposite notions. An experience in this partially undefined and contradictory in-between state both allows us and forces us to consider new ways of being—to transform. In liminality, we know that what we see is not our existing world, yet we experience an immediate feeling of something very real. This creates a need for accommodation between these two levels of "reality." Think, for example, of an adolescent who is in a transformative state from a child to an adult identity, which requires them to explore a new understanding of their personal reality (see Larson et al., 1996). If we venture too far into the undefined, it becomes meaningless, random, and chaotic, not to mention terrifying. Stay too close to what is familiar and safe, and we become bored, calcified, and unable to adapt to change. This ties liminality directly to narrative structures: the idea that all stories need to be "inevitable yet surprising" at the same time.

We rationalize sensory input by creating causal models of the world around us, how it works, why and - based on that - what might happen next. These are narrative models. Putting on hold some beliefs from our existing narrative and entering a liminal space allows us to safely test, play with and consider alternative narratives (Gaggioli, 2016; Kitson et al., 2019). It is this process of re-evaluating and changing one's own narrative of the world and self that is the essence of personal transformation. We can think of liminal design as a version of the Cartesian notion of the pineal gland[1]: a stage to imagine the self, and to create narrative meaning out of the sensory inputs from the world around us.

Are all good user experience designs liminal? No, not all designs require liminality, nor do they necessarily seek transformation of any kind. In this context, the goal of Liminal Design is to provide principles to guide the design of experiences that help individuals find new meaning, enhance their emotional and moral abilities, increase inspiration, creativity and imagination and support transcendence. Designing liminal spaces also means helping people to explore new "spaces of the self".

---

[1] Descartes regarded the pineal gland, a tiny organ in the center of the brain, as the principal seat of the human soul (see, Abhyankar, R. (2020).

The goal of most traditional UX is to increase usability, pleasure, and satisfaction (Norman, 2004), whereas Liminal Design tries to go beyond these pragmatic and hedonic aspects and explicitly addresses the eudaimonic sphere of wellbeing (i.e., autonomy, meaning, purpose, self-actualization, and transcendence, see Ryan & Deci, 2001). Clearly, when designing liminal experiences, one needs to work at a higher level of abstraction, which reduces the possibility of directly translating "user needs" into "design requirements". Liminal Design is, in this respect, more like the process of artistic creation than conventional design. Consequently, this highly dynamic context requires us to look well past simple technological fixes to create meaning, and instead consider problems and solutions as meaningful social situations created through interplay between the narrative expectations we arrive with and the design we experience through participation.

This contribution is structured in four sections. The first gives a short overview of previous work that explored how design can stimulate and facilitate experiences that extend beyond simple pleasure or satisfaction. In the second, we provide a short look at the seminal work that has come before us connecting transformation and liminality and explain the motivation and background for our framework. In the third, we outline the foundational assumptions of Liminal Design and a practical design model illustrated by examples and potential applications. In the fourth and last section, we discuss higher-level implications of Liminal Design.

**Previous Work on Design Beyond Pleasure and Efficiency**

Traditional UX design focuses on usability, pleasure, and satisfaction. Various design researchers have challenged this dominant perspective and explored how the UX repertoire can be enriched beyond the purely utilitarian and hedonic. To start, several researchers have proposed that the traditional focus on generalized pleasure in design research does injustice to the differentiated nature of human emotion. Design can evoke a diverse palette of distinct (positive) emotions, for example, hope, pride, fascination, relief, or love (Desmet, 2002). Although all positive, these emotions are essentially different – both in terms of the conditions that elicit them and in terms of their effects on human-product interaction. For example, whereas fascination encourages a focused interaction, joy encourages an interaction that is playful (Fredrickson & Cohn, 2008). Various frameworks have been introduced to support this more granular understanding of pleasure. Desmet (2012) introduced a typology of 25 positive emotions that provides a fine-grained yet concise vocabulary of positive emotions in human-product interactions. More recently, this typology was developed further into a detailed online emotion database (https://emotiontypology.com). This typology includes complex experiences that can include transformative qualities, such as elevation, serenity, and awe. Yoon, Pohlmeyer, and Desmet (2016) introduced four design tools that can help leverage this differentiated nature of positive emotions in design processes. One of these tools is a card deck: the "Positive Emotional Granularity Cards" that aims to increase the designers' so-called *positive emotional granularity*, which reflects the degree to which a person can represent positive emotions with precision and specificity (Tugade, Fredrickson, & Feldman Barrett, 2004).

Other authors have gone even further in scrutinizing pleasure in design usage by exploring how negative emotions can make a meaningful contribution to user experiences. Fokkinga and Desmet (2012, 2013) proposed that designers can enrich user experiences by purposefully involving negative emotions in user-product interaction. They introduced a framework of rich experiences, which explains how and under what circumstances negative emotions make a product experience richer and even more enjoyable. Through several design explorations, they demonstrated that negative emotions are a viable and interesting starting point for creating emotionally rich product experiences. This proposition aligns with the work of various authors who explored how product experiences can more truthfully mirror the richness of real-life experiences, and even enrich and expand our experience of everyday life. Hassenzahl (2010) suggested that product experiences should be "worthwhile" or "valuable" to avoid the pitfall of aiming for shallow pleasure in experience design. Likewise, Arrasvuori et al. (2010) investigated the possibilities to create more engaging consumer products by using the wide range of emotions that people typically experience when playing video games. With their concept of "design noir,"

Dunne and Raby (2001, p. 45) even proposed a new genre of design to complement the prevailing "Hollywood" tradition of products that offer a limited experience.

Similar to the critique on basic pleasure, various authors have challenged the narrow operationalization of usability as a measure of effectiveness and efficiency, especially in behavioral design (i.e., design that aims to stimulate altitudinal and behavioral change). Laschke, Diefenbach, and Hassenzahl (2015) introduced a design approach in which "situated friction" plays a central role (see also Laschke et al., 2014; Hassenzahl & Laschke, 2014; De Haan, 2021). This approach creates frictional feedback to disrupt routines and stimulate people to imply alternative courses of action. By deliberately reducing usage efficiency, friction inspires reflection and meaning making. The authors present several design cases that explore how frictional feedback can be experienced as acceptable and meaningful, while stimulating the intended reflection. In a similar fashion, Rozendaal et al. (2011), and Boon, Rozendaal, and Stappers (2018) introduced a design perspective that emphasizes ambiguity (i.e., affording multiple interpretations) and open-endedness (i.e., affording multiple courses of action). While design ambiguity and open-endedness typically reduce usability in a traditional view, these authors explored how these qualities can contribute to user experience, leaving room for autonomy and creativity of end users in solving problems, creating meaning, and determining product usage.

As a final note, we should mention that several authors have started to explore how technology can be used to create experiences that can be described as spiritual or *transcendent* (for a recent review, see Blythe and Buie, 2021; Buie, 2018). While diverse, all these inquiries focus on design for a sense of openness and unity — experiences of connection with something that is larger and more permanent than oneself, which comes with emotional experiences that are deeper, more impactful, and more profound that simple pleasure, such as experiences of ecstasy, tranquility, gratitude, awe, and reverence. To explore transcendence design, researchers have experimented with design spaces that are ineffable, ephemeral, or paradoxical.

These initiatives exemplify an increasing awareness in design research and practice that a mere focus on basic usability, pleasure, and satisfaction represents design intentions that are inherently narrow, and that knowledge and methodology are required that enable and support the pursuit of design that better represents the infinitely complex and rich repertoire of human experience. On a more general level, these initiatives recognize a need for 'alternative ways of knowing' to compliment today's apparent ethos of consumerism and materialism that is stimulated by the high rates of scientific and technological progress. They explore a new space for the development of a material culture that is in greater harmony with one's inner development and their outer morality and (self)compassion (for a discussion, see Walker, 2013). A current challenge is that little guidance is available on *how* to design for this kind of transformative design. With our Liminal Design model, we aim to contribute to this progress by providing a structured approach to transformative design that provides the user with a safe space for their personal and complex inquiry towards transitional development through value formation, interconnectedness and feelings of unity.

**Foundations for Liminal Design**

Many of us multitask our way through activities traditionally designed for a separate and committed time and space, with their own rules, hopes and expectations. Wildly different types of activities and tasks are streamlined onto a single technological platform, leaving disparate experiences mashed up on a single screen filtered through a single branded aesthetic and interface - all to be found in the very same geographic location as the rest of our daily routine and no longer requiring the physical movement that once helped distinguish between them. A contemporary example is the increased use of video conferencing tools that arguably has left us drained (Fauville et al., 2021; Wiederhold, 2020) and at times thinking that a simple phone call in a quiet place might deliver more intimacy. Liminality has been pushed aside in favor of platform efficiency. The sublime state is struggling to find room in a digital reality built for commerce.

However, this situation also presents an important opportunity for design to take back what we have lost and to push human experiences into new, deeper, and more interesting realms. To further contextualize the need for a shift in design approach, we should also consider macro shifts in the economy: from an industrial core in the 1800s, to information-based in the 1970s, and more recently to the so-called "experience and transformation economy"(Pine B.J. and Gilmore, 2019; Pine and Gilmore, 1999). Despite this latest advance, tools, methods, and technologies have struggled to manifest its theoretical promise (Gaggioli, 2016). We are still using outdated industrial and information models to solve a new era of experience problems. The big technology companies are struggling with this shift for similar reasons: their models rely on enormous scale, uniformity, and predictability to support a particular commercial driver, a financial platform often decoupled from the core value proposition of the technology delivered.

We are, however, not interested in piling more of the same onto the now-familiar tech skepticism stemming from writings such as "The Filter Bubble" (Pariser, 2011) and "The Age of Surveillance Capitalism" (Zuboff, 2019). Instead, a new practical design framework is urgently needed: not to look for incremental improvements within established product strategies, but to explore and develop radically new ones.

In this spirit, the Liminal Design model we propose always starts with *first principles* to define the right problem that we should focus on solving. Only then can it establish a practical process to highlight and sequentially organize quintessential concepts of liminality, transcendence, and transformation in the shape of new products. This process will always need to first consider the *why* as the foundation for the narratives that we invite users to. Without a narrative that we can care about and can want to make a personal investment in, no technology can deliver what we need for liminality.

Winnicott (Winnicott, 1971) described a "potential space" as a metaphorical expanse that is intermediate between fantasy and reality, an area of experiencing which opens new possibilities for imagination, symbolization, and creativity. According to Winnicott, potential space is inhabited by play. "It is in playing and only in playing that the individual child or adult is able to be creative and to use the whole personality, and it is only in being creative that the individual discovers the self" (p. 54).

The notion of liminality (from the Latin term *limen*: threshold, boundary) was first introduced by the ethnologist Arnold van Gennep (van Gennep, 1960) to describe the initiation rites of young members of a tribe, which fall into three structural phases: separation, transition, and incorporation. Van Gennep defined the middle stage in a rite of passage (transition) as a "liminal period". Elaborating on van Gennep's work, anthropologist Victor Turner (Turner, 1974, 1981) argued that, in postindustrial societies, traditional rites of passage had lost much of their importance and have been progressively replaced by "liminoid" spaces. They are defined by Turner as "out-of-the-ordinary" experiences set aside from productive labor. These liminoid spaces have similar functions and characteristics as liminal spaces, disorienting the individual from everyday routines and habits and situating him or her in new circumstances and narratives that deconstruct the "meaningfulness of ordinary life" (Turner & Turner, 1985, p. 160).

The metaphors of potential space and liminality/liminoid space provide a platform for further elaborating the purpose of transformative design as the realization of interactive systems that allow participants to experience generative moments of change. However, as open-ended "experiments of the self", such interactive, transformative experience may also situate the participants in situations of discomfort, disorientation, and puzzlement, which are turning points out of which new possibilities arise (Gaggioli, 2016; Kitson et al., 2019; Riva et al., 2016).

The Liminal Design model is built around four foundational assumptions, illustrated in Figure 1:

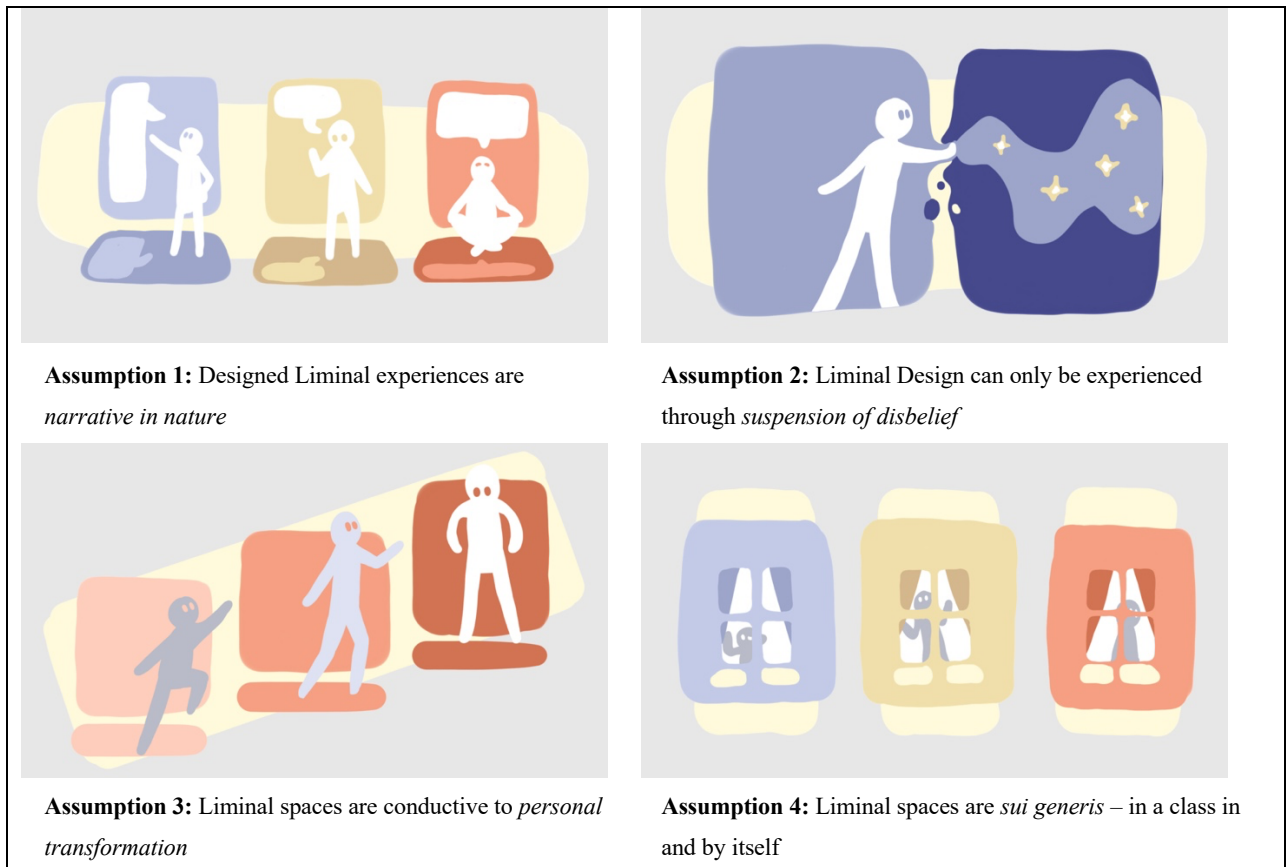

**Figure 1.** Four foundational assumptions of the Liminal Design Model

Assumption 1: Designed Liminal experiences are *narrative in nature*.

The purpose of liminality is the creation of a space within which we are allowed to play with - and consider - new narratives of the world and of ourselves within it. Humans organize experiences and sensemaking in a narrative form (Brockmeier, 1995, 2009; Bruner, 1991) that requires semiotic distinctions or borders to be meaningful (de Luca Picione and Valsiner, 2017): me versus not-me, here versus there, past versus future. The liminal space is a semipermeable reality between where we have enough stability to retain a sense of self without threat of chaos and disintegration, and - at the same time - enough openness to the un-narrated outside and its endless possibilities. The ambiguity between the two calls for us to accommodate by creating our own new narrative to better hold the contradiction. For designed liminal spaces - in contrast to general awe or external catastrophic disruption - the dialogue suggested between one reality and the other is both targeted and specific, i.e., it offers an experience that suggests a certain narrative path of personal transformation. De Luca Picione and Valsinerb beautifully explore this complex interplay between semiotic structure and narrative liminality (de Luca Picione and Valsiner, 2017):

"The peculiar dynamics and the semiotic structure of borders generate a liminal space, which is characterized by instability, by a blurred space-time distinction and by ambiguities in the semantic and syntactic processes of sensemaking. The psychological processes that occur in liminal space are strongly affectively loaded, yet it is exactly the setting and activation of liminality processes that lead to novelty and creativity and enable the creation of new narrative forms." (p. 1)

Assumption 2: Liminal Design can only be experienced through *suspension of disbelief*.

The space that Liminal Design suggests is by nature overtly different from our day-to-day reality. It is a construction to be experienced for a limited period for the explicit purpose of creating liminality within a specific narrative. To play a part and experience anything in it, we must first decide to enter it and be open to leaving some assumptions about ourselves and our world behind for a moment: to suspend disbelief. Instead of simply rejecting the liminal construction as untrue or unreal, we are asked to play with - and to have first-hand experiences of - its suggested possibilities.

The success of suspension of disbelief is derived from a dialectical relationship between our desire to experience something new and a safe but immersive-enough violation of expectations inside the liminal space to confirm that we are closer to something intangible that we desire. This is the narrative continuously unfolding and being made as we stay in the liminal space. To allow oneself to be lost in the highly constructed universe of a book or a film is analogous to this experience, as is any profound experience of art (see Schaper, 1978). The true beauty of suspension of disbelief is that of being freed from the limitations of "reality" and therefore open to a glimpse of what is infinitely bigger than our own narratives: the sublime.

The famous quote from Samuel Taylor Coleridge, who first coined the term *suspension of disbelief* (*Biographia Literaria*, Chapter XIV, 1817), certainly captures the promise of what we have in mind:

"...to give the charm of novelty to things of every day, and to excite a feeling analogous to the supernatural, by awakening the mind's attention from the lethargy of custom, and directing it to the loveliness and the wonders of the world before us."

Assumption 3: Liminal spaces are conducive to *personal transformation*.

The very present and yet un-real experience of a designed liminal environment challenges our assumptions and semiotic borders and thus questions our canonical narratives of self. To reconcile this unreal narrative universe with what we so unmistakably feel are very real emotions, we consider and play with new personal narratives.

So, as we exit the liminal space and return to our regular reality, the most basic hope is that the experience and new ideas generated within still resonate, thus allowing us to see our old world in a slightly new light and with more agency to choose how we play a part in it. The experience has transformed us (Gaggioli, 2020).

Assumption 4: Liminal spaces are *sui generis* – in a class in and by itself.

All liminal spaces are sui generis, meaning that they are a class in and by itself that extends beyond conventional genre boundaries. This implies that they cannot be approached or understood outside of - or as part of - any other structure or hierarchy than themselves. It is precisely their lack of attachment to practical functionality or existing taxonomy that creates their liminality, and thus allows room to experience a violation of the expected. Because we are interested in designing liminality to promote personal transformation, it is specifically the semiotic ambiguity between the profane and the sublime that we exploit for purposes of a given narrative, i.e., that narrative is not part of an already existing construct. The liminal space experienced, the space between two opposites, is always experienced in its entirety, not as a smaller part of something established. This makes it sui generis.

"This interstitial passage between fixed identifications opens up the possibility of a cultural hybridity that entertains difference without an assumed or imposed hierarchy" (Bhabha, 1994)

In addition, we must consider *presence* - our participation in the liminal - as sui generis as well. The singularity of a liminal experience and its ability to transform the self can only be found in the dialectical play between opposites: *me versus not-me, here versus there and past versus future* (de Luca Picione and Valsiner, 2017; Gaggioli, 2016; Marsico, 2011, 2016).

**An approach to Liminal Design**

The practical approach to Liminal Design consists of three sequential steps (see Figure 2): Narrative Desire (i.e., selecting and building a narrative stage), Optimized Abstraction (i.e., optimizing the space for the targeted experience), and Suspension of Disbelief (i.e., creating desired participation from start to finish). By reinfusing created experiences with these elements, the design opens existing spaces for imaginative and even transcendent engagement with the world and with others. The goal is not a faithful restoration of analogue physical experiences, but rather a plumbing of new types of interpersonal contact and imaginative experience not possible prior to today's technology.

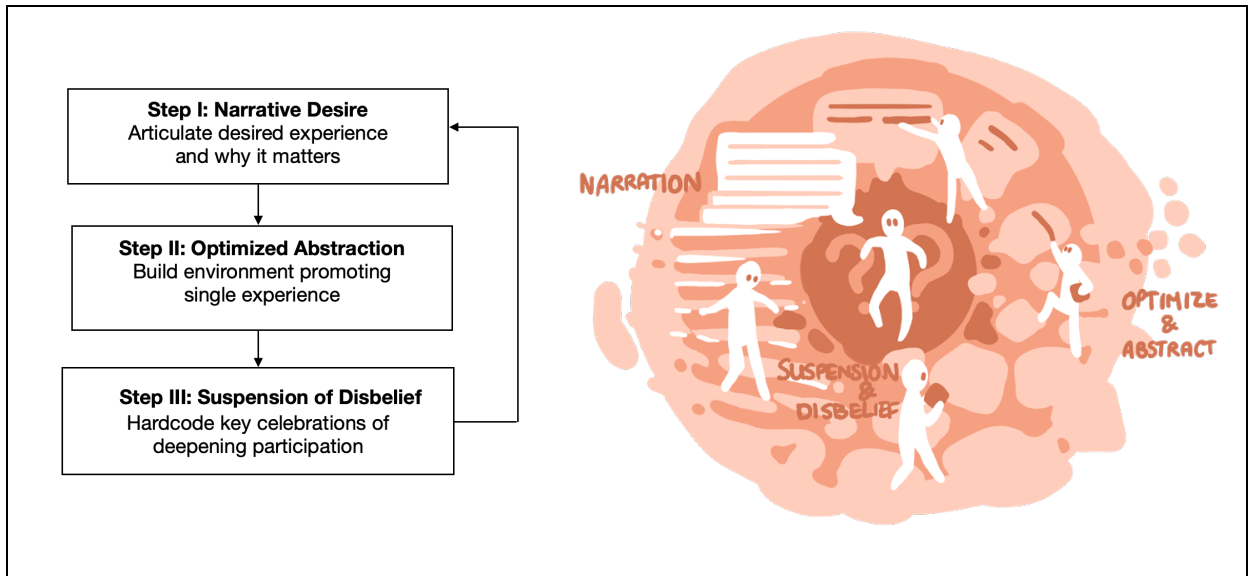

Figure 2. Three-step approach to design for liminal experiences

The Liminal Design approach is not intended as a formula to mechanically follow. As with all creative endeavors, good solutions will require a committed leap from strategy to the specifics of implementation and manifestation in the design. A technological design change cannot be considered in isolation, but only has meaning in the context of the dynamic interplay between narrative and participation, mirroring our three steps below (Dorst & Cross, 2001).

Step one - Narrative Desire

Narrative Desire is the articulation of the expectations and attitudes we want users to approach our experience with. There are many ways that we can set this context: marketing and PR, packaging, stories, instructions, and product mythology. All of these can prepare the user to understand and navigate an abstract liminal space. At the same time - and with the same message - this context must also create a real desire to participate and seek the intangible of what can only be had inside the liminality offered up by the product.

As we discuss in step two below, abstraction is essential to designing and experiencing liminal spaces. Narrative Desire gives the collective abstractions purpose and direction. When the product holding the liminal space is designed properly, every step in the experience leads us deeper towards the desired transcendent target. It should be noted, as with all narrative and story, that although each individual part of the experience cannot - and should not - hold the entire narrative, all parts need to have a role and a function in an efficient, singular, and coherent overarching narrative.

It cannot be stressed enough that for all stages of Liminal Design, it is imperative that we choose a Narrative Desire that is important enough to care about. As with all stories, if we don't care about the narrative we will not invest in the experience and never make the commitment necessary to liminality, transcendence, and transformation.

Step two - Optimized Abstraction

Once the Narrative Desire is clear, two key components guide the design of a liminal space to manifest the narrative: how we *optimize* and *abstract* a suitable stage for a user's transcendence that promotes the specific target experience.

*Optimization* and *Abstraction* are coupled here because each is embodied through the other. We *optimize* by deleting what we don't need, and thereby keep front and center what we want the experience to focus on. This purposeful reduction of everyday reality creates *abstraction*. In reducing the distractions of the environment to highlight and promote a specific singular experience, Abstraction becomes Optimization. When we participate and have experiences in abstractions, we experience a liminal space.

To *Optimize*, we deconstruct the desired narrative experience: its quintessential elements and their sequence. It is important to note that the design does *not* deliver the desired experience. It simply holds the user in an optimal environment for the targeted experience to be had. For example, focus is typically required for immersion: an immersive experience should therefore provide the right single focal point and design the rest of the space to eliminate all distractions, physical and digital.

*Abstraction* - understood separately from Optimization - creates room for us to bring and project a narrative onto an experience. Even if objects in our designed experience are familiar, they will still be curated or designed objects, functioning as tokens, visuals, and icons (Mitchell, 2015) that are separated from their usual reality and context. In other words, they are abstracted. They will then have to be filled with new meaning, within the context of the liminal space, by the user projecting themselves into a version of the Desired Narrative that is meaningful to them. Religious practices and places of worship often use and then fill ordinary objects with story and grant them sacred status as part of creating a liminal place.

Our work is directly related: we are attaching meaning to objects needed to play out an alternative narrative experience. We are not replicating reality. We are creating a new one. This is part of the abstraction and will therefore, as part of how it is constructed and experienced, be sui generis in nature.

If an object in our space is still part of its traditional and contextual reality - a button to turn the experience off or adjust the volume, etc. - we would probably be well served to make sure it doesn't contradict or distract from the desired narrative. Keeping non-abstract functional objects out of sight is often the best solution to keep non-narrative stimuli to a minimum. And if we can't hide them, there are often ways to visualize and ceremonialize these functions and actions to turn them into tokens of meaning in the liminal narrative context.

Now, all narratives experience changes and morph through their arcs: no experience should feel the same at the beginning as it does at the end. Optimizing for the arc supports different targeted changes throughout the experience. These changes anticipate and manifest key aspects of the targeted narrative experience, much like the music score of a film helps amplify the experience the filmmaker wants us to have and cue where they want to transport us next. A deep understanding of the behavioral transitions between early and late stages of the experience is imperative. This way the design can support and hold each step while creating enough desire for the user to seek the deepening of the experience coming up next. To do this we need to maintain a user's presence in the liminal space and its narrative - and for this we have the help of the third step in the Liminal Design approach: *Suspension of Disbelief*.

Step three – Suspension of Disbelief

All the experiences we have discussed are personal to the user and intentionally encountered by them. But no matter how well we choose a Narrative Desire, or how perfectly we optimize the abstraction of our stage for the experience, there's no causal effect that we can rely on to trigger transcendence. *Suspension of Disbelief* can, however, be promoted at key points in the experience arc: first by establishing an *independent space* for liminality, then by *ritualizing* and *celebrating* targeted behavior through the experience arch. And lastly by allowing enough *narrative room* in the experience to afford users flexibility to make the experience their own. These three aspects all support immersion into - and throughout - the liminal space and its narrative, i.e., suspension of disbelief.

Independent Space - Physically and spatially dedicating a space to the target experience is the most straightforward solution to designing a liminal space. Even the earliest settlements, created as far back as 15,000 years ago in Turkey, show humans distinguishing between everyday spaces and spaces dedicated to worship. An optimized space is not liminal until we have clearly defined its perimeter as different from that of our ordinary reality. A space can be both external (a theater, church, tearoom, or nightclub, etc.) and internal (closing our eyes before sleep, meditating, praying, taking psychedelic drugs, listening to a story etc.), but in all cases it relies on a semiotic distinction to create the very border we are to traverse.

Research shows that imagination and memory are in part spatially organized (Robin, 2018). The mnemonic device "Memory Palace" (Yates, 1966) is one example of this. Consequently, our minds have evolved to partly reset as we move between defined spaces. "The Doorway Effect" is a common phenomenon where we might walk into the kitchen only to forget why we did so (Radvansky et al., 2010). Since our goal is immersion in the liminal space, clarity to that moment when we cross from one space to another aids suspension of disbelief: demanding a decision and commitment from us to either stay in our ordinary reality *or* to step across the border and into the liminal journey. The latter choice relies on us pushing aside the noise and concerns of the existing world and being ready for something new and different. It is often meaningful to think about Liminal Design as the creation of an attitude conducive to transformation. It again speaks to the dynamic nature of this task: the continuous dialectical play between user, space, and narrative.

Ceremony - In this context, *Ceremonies* are targeted celebrations of steps that support the underlying sequential narrative helping us to stay present in, and to move us from the beginning to the end of the liminal experience by. What tactile, aural, and visual feedback can we design to manifest mile-markers of where we are in the narrative arc, and more importantly, to highlight that narrative to create a desire to continue.

Ceremony can be an agreed-upon ritual: applause and cheers between songs at a rock concert, a sports team's huddle before the game, a writer clearing the desk of clutter before a day of writing. It can also be hard-coded into the design, content, or UX. "Intro titles" (or main titles) for films and television are one very direct example of Liminal Design: they leverage the required disclosure of production credits in order to bring a "fresh audience" into the specific fictional - or liminal - universe of a film or television series using abstracted audio-visual and narrative cues. Another example is intentional phenomenology in architecture: how a building's design deliberately provides different choreographed experiences and feedback depending on place, angle of view, time of day, etc. An interactive UX might change the soundscape to protect and gently anticipate the upcoming targeted experience, putting it into clearer focus as an act of creating desire.

Narrative Room - Lastly, it can be tempting to be prescriptive in articulating Narrative Desire. Clarity about why the experience is to be desired is important, but a too tightly held notion of *how* a user must experience every detail of it will run the risk of not leaving enough *Narrative Room* for the individual user to make the journey their own. As happens with all stories, we make it relevant to us by projecting our own hopes and fears into the narrative. The design of the story must provide room for these projections to be made.

This is even more true with liminal spaces. The bigger and more complex emotions that we are designing for can run a higher psychological risk in letting go of one reality and losing control of the familiar. Transcendence is very personal, as are the psychological hurdles we must clear to participate fully. Therefore, all liminal experiences are abstract. Transcendence is and must be a personal experience of becoming part of something greater. Conversely, when abstraction is replaced with detailed instruction, there is no longer space for us to imagine ourselves in a different reality and to participate through projection and on our own terms.

It is only when the imagination is stirred - giving us a glimpse and hope of what we desire - that Suspension of Disbelief is likely to take place. And when it does, liminality can offer up personal experiences far more intense and sublime than what is merely prescribed. Striking the right balance between guiding the targeted narrative desire and offering enough narrative openness for the user through abstractions is the core creative work of Liminal Design.

With the proper Narrative Room, users of liminal spaces can also show a very useful opportunism in what they choose to incorporate into the narrative and how. For example, the active suspension of disbelief during a video conferencing call might include users ignoring the common knowledge that their screen just presents a mediated digital image of someone real far away. At the same time, to mitigate their shyness or hesitation about intimacy, they might simultaneously hold a contradictory fact as part of their narrative: that the other user is thousands of miles away, thus making the experience less threatening. Only with enough Narrative Room and a motivated user will we unlock the unique power of liminality: to create heightened experiences and personal transformation. In addition, a liminal experience - when explored opportunistically and beyond the limits or reality - has the potential to be more profound and more intense than any real-world alternative.

For all aspects of the Liminal Design model, it should be emphasized that liminality per definition must - contrary to the goal of most other UX work - avoid smooth, routine, or transactional relationships with the user. Designing for liminality is always an effort to encourage focus and renew deep commitment and investment in interesting and meaningful experiences. The latter is done by making a user's participation a clear choice and then manifesting its meaning through an intentional narrative. And we can only know that narrative from the friction we experience - and the hurdles that we chose to clear - because of it. There is no story without conflict.

**Liminal Design in practice: Remote conversations**

To illustrate our Liminal Design model and the three related design steps with a concrete example, let's go back to our earlier mention of connecting through video conferencing (VC). We explored an alternative design solution for video conferencing; one that permits the experience of deep presence and connectedness. Figure 3 is a visual representation of the conceptual design space that emerged in this process. Below we describe step-by-step how this design space was created.

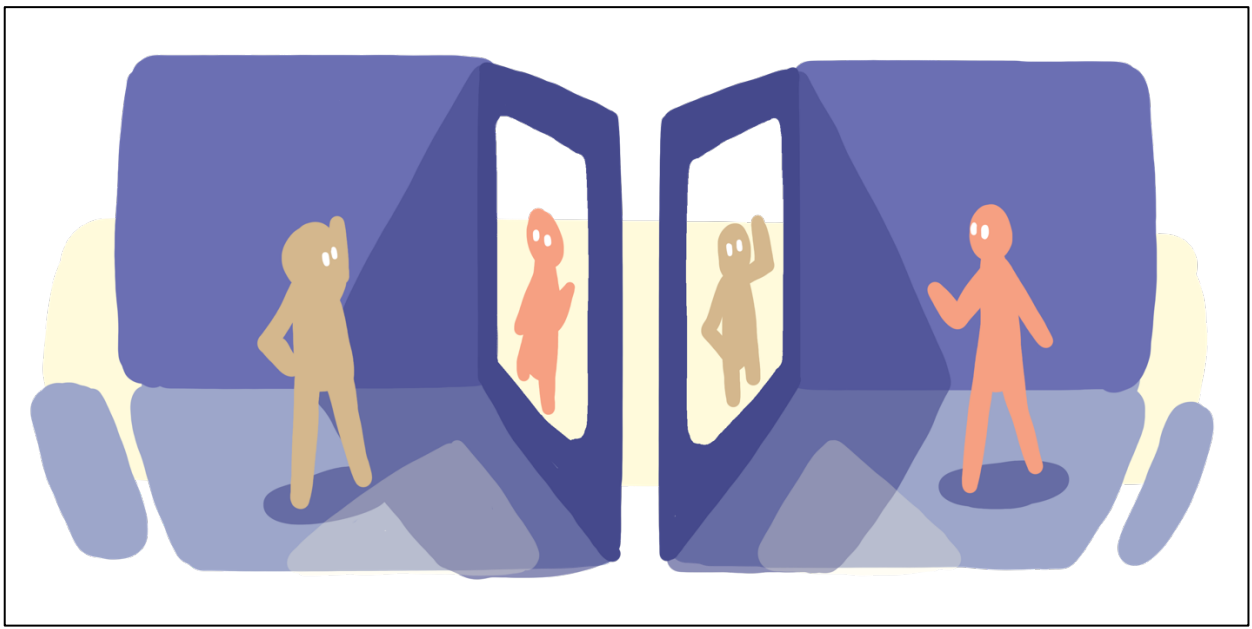

**Figure 3:** Liminal Design in practice: A space for deep presence in online conversations

Although using ubiquitous applications such as Zoom, Microsoft Teams or Google Meet offers an almost live image of other participants, its increase in use over the last few years has highlighted many of its shortcomings in providing real social connection, as well as the well-documented exhaustion often referred to as "Zoom fatigue" (Aagaard, 2022; Bailenson, 2021). Let's look at some of the existing research and explore how liminality might offer new ways to frame up the problem space.

Since the seminal work of Albert Mehrabian (Mehrabian, 1972) in the late 1960s, we know that a significant part of human communication is carried out through non-verbal cues. In any face-to-face interaction between two people, both participants are continuously exchanging a variety of social signals (i.e., gestures, facial expressions, posture, body orientation, etc.). To achieve successful communication, we must process incoming signals and send back meaningful signals at a suitable pace. Indeed, the lack of nonverbal cues has been proposed by early theories as a causal factor in accounting for the (in general negative) differences between computer-mediated and face-to-face communication (Daft and Lengel, 1986; Kiesler et al., 1984; Short et al., 1976). The lack of visible non-verbal cues in most VC mediation, it has been argued, will not simply have us pause our instinct to try to read others as we have been taught by thousands of years of evolution, but will instead have us continuously *try harder* to read what is not available in the low-fidelity image, causing higher cognitive load and fatigue. The work of Hiroshi Ishii (Ishii and Ullmer, 1997), head of the Tangible Media Group at MIT, is also relevant here: suggesting that our highly evolved tactile skills are yet another missing piece in most digitally mediated communication.

However, other models suggest that an *excess* of non-verbal cues could be just as deleterious to an effective mediated communication experience. For example, in examining the potential psychological causes of so-called "Zoom fatigue" Bailenson (Bailenson, 2021) proposed the main causes to be: excess of eye gaze at close distance (to the screen); disproportionate size of head compared to regular field of view; unnatural body positions, as we need to stay still in camera view to be seen; and gestures needing to be exaggerated to be understood and not always being the same as in a face-to-face interaction.

Some research would take all this even further and argue that our most common VC applications, from a social perspective, are likely to do more harm than good. For example, although the lack of direct eye contact has a very clear explanation with the camera placed above our gaze on the screen, our subconscious will still tell us that the person not looking us in our eyes is trying to deceive us (Bekkering and Shim, 2006), eroding both trust and connection.

Whether we get too much or too little information, overwhelming evidence suggests that our most-used VC tools come up short in delivering even the most basic social functions of in-person conversations. That said - and this is key in how we must approach Liminal Design – the corporate context in which most VC calls take place, even in-person conversations are likely to leave us wanting more in terms of deeper discussions, purpose, intimacy, and connection. This is the main reason the United States and other developed countries are experiencing an unprecedented rate of employees leaving their jobs, a phenomenon recently referred to as "The Great Resignation" (U.S. workers give employers high marks for supporting them during the pandemic - WTW, n.d.). Simply increasing the resolution of VC applications will not solve this.

Therefore, the first task for Liminal Design is to better align a solution with a narrative that participants are willing to invest in. As with any book or film, if the story and the stakes aren't important, we will simply not invest in a narrative enough to make it our own. Without participation in a narrative, abstraction with meaning and liminality will be absent, thus never presenting the user with an opportunity for transformation. There is no Liminal Design without an ambitious narrative. This is where we must start.

Step one: Narrative Desire

Our focus here is not just any solution, but one that achieves liminality. As we think about connecting two people, it therefore seems natural that we skip past transactional or casual chats to instead look closer at conversations that are deep and intimate. Think of two people fully immersed in conversation at a busy restaurant: leaning in over the table, completely focused on each other and allowing space, time and reality around them to fade away. Together they create a singular interconnectedness in a real-time shared space - an experience that psychologists define as "group flow". The social and psychological benefits of such experience have been well researched: empathy, trust, and creativity, to mention a few (Gaggioli et al., 2011, 2013; Sawyer, 2006).

Given this problem space, a good Narrative Desire for our design can be *a liminal space to inspire, hold and engage two people thousands of miles apart in important conversations leading to interconnectedness as deep as - or deeper than - in-person meetings*.

It is important to note here that the articulated Narrative Desire is not for *all conversations*, but for *important conversations.* This clearly delineates our effort from the many times Zoom is used for transactional exchanges where a simple email might suffice, thus not requiring any liminality to work. This is the foundation of Narrative Desire: to clearly align the design with a meaningful narrative target that can only be found on the other side of some level of transcendence. This perspective of a meaning created in a dialectical relationship between the design and Narrative Desire implicitly argues that our issues with Zoom might be addressed, not by changing the product, but instead by changing the narrative that we contextualize and use the product within.

From a corporate perspective, this is a big and important problem to solve in a world adjusting to remote work, increased focus on the environmental cost of business travel and long-term trends continually pushing towards more internationally distributed work groups. The articulation of Narrative Desire is a foundation to guide the design work while simultaneously creating inspiration and motivation for the user to participate in the experience.

Step two: Optimized Abstraction

Let's go back and look closer at the scenario we are solving: two people absorbed in a conversation and the deeper presence associated with it. What elements are required, and to what extent can we replicate them in a remote and mediated setting? There is existing research available to inform both - four critical elements often mentioned are:

1. Direct eye-gaze. This we can easily provide with a one-way mirror setup akin to how a teleprompter works.
2. Non-verbal cues. High-resolution camera, screen for audio and video and a low-latency network would likely do a good job of communicating more fidelity than most in-person settings. Ideally, we should be able to read more than just the face and crop the image of both participants just below the hips, including gesturing and resting hands.
3. Natural scale and distance. If we use a large enough screen (55 inches) to render part of a human at natural size, placed at natural conversational distance from the other participant (about 4 feet), this aspect should be covered. It should also be sufficient to handle non-verbal cues per above.
4. Shared sense of space. As we are designing a remote and mediated setting, this requirement is different: we would have to rely on illusion rather than a functional solution. We could again borrow from the established world of theater and cinemas: to hold the experience in darkness on both ends and only portrait-light the participants. This would hide the technological mediation and allow participants to fully focus on each other during the conversation in what appears to be a shared space of darkness.

This prototype suggesting the shape of a cinematic VC booth embodies both *abstraction* and *optimization*. The space provides the necessary ingredients for in-depth conversations as well as an almost complete reduction of all things that might distract from the same. As a separate space - say a black painted booth dedicated to important conversations, nothing else - in an otherwise regular office environment - the abstraction is materialized through its own semiotic distinction coupled with the expectation that deep connection will be provided despite mediation. Through use and participation, we agree to be part of and to help create the illusion.

Step three: Suspension of Disbelief

This last part of *agreeing to be part of the illusion* is Suspension of Disbelief. What might we do and design into the experience to encourage it, hold it and hopefully encourage an arc of deepening participation in the narrative? Let's explore this through the lens of our Liminal Design model:

Dedicated Space - With the above notion of a black remote conversation booth, the dedicated space is already well articulated. We can add to this by ensuring colors, style and materials are different from the corporate normative pallet. In addition, marketing, instructions, PR as well as icons and UX would stay clear of vernacular used with traditional VC applications ("meeting", "call") and instead emphasize the more intimate universe and narrative desire it is setting up: *the corporate meeting is dead, long live important professional conversations.*

Ceremony - Consistent with the idea of reducing noise and stimuli that don't support the focus on the other and immersion in the space, we might ask participants to leave devices such as mobile phones and computers outside. This is likely to feel unnatural to many, but, as discussed above, it is this friction and hurdle that commits us to a different experience. We are ritualizing part of the optimized abstraction and literally manifesting the shedding of our ordinary universe to step across the threshold to another one.

It might also be that our prototype could be further enhanced to nudge the experience by gradually cropping the image (zooming in) - moving imperceptibly closer - as the conversation progresses, and at the same time having the voices render deeper and the lighting gradually move to a lower color temperature. The aim would be to mimic a real-world conversation of two people moving closer as they get more interconnected. The warmer light would signal intimacy. This, we argue - if done delicately without breaking the trust in the experience - is likely to have participants act accordingly and reinforce the arc of Narrative Desire of a successful and deep conversation.

Narrative Room - The suggested product design and the articulation of Narrative Desire envisioned here does not run the risk of being too prescriptive or exclusive. This is proportionally important to the elements of friction we have designed, which require a heightened commitment to an experience that is potentially both highly intimate and intense. For example, if you are shy, you might choose to keep in mind the actual geographic distance to the person you are connecting with in another country - not the distance to the screen you are watching that creates the illusion of someone's nearby presence. You opportunistically navigate and fill out the abstraction of our product's dark environment to support the desired narrative and, if needed, your highly individual needs for participating fully. Similarly, it is worth noting for this example that this sound-proof, dark, optimized, and abstract space has the potential of producing a higher degree of presence and interconnectedness than a real-world environment: a hyper presence of sorts.

**Other application examples**

As we quickly touch on each example below, it should be clear that any solution offered is but one example of how we might approach and leverage liminality to target a specific narrative goal. It is not the *only one,* or the *right* one - it is just one theoretical example of the process applied.

*A Restaurant*
A restaurant is in many respects already a separate and liminal space: clear separation from other services and places, enforced through rules of social conduct and legislation. We must decide what aspect of the restaurant we want to tackle and make our Narrative Desire problem statement, and then explore how we might make that aspect more interesting through liminality. We would of course make very different design choices if we wanted to deepen social interaction, highlight wine selection, enhance the bar's dating scene, or call attention to the ethnicity of the cuisine. Let's pick one of the most basic: elevating the taste experience of food.

One possible approach to deliberately optimize and abstract, would be to strip away one whole category of sensory stimuli, such as sight and host the dining experience in complete darkness. It's a significant and deliberate violation of routines and habits around dining presented as safe for exploration. Liminality is offered. Because the space it is labeled restaurant, albeit slightly different, the narrative desire is clear even if implicit. As guests, we are now allowed - perhaps even forced - to focus more attention on the sensory inputs left: in particular, taste. In addition, because we cannot see the food that we eat, there is an additional heightened sense of anticipation and risk that plays up the personal commitment necessary to activate this particular liminality. Based on the 1999 Blindekuh restaurant in Zurich, there are now an estimated dozen

"dark dining experiences" around the world, including CamaJe Bistro in New York, the three Opaque restaurants in California as well as Dans le Noir in Paris[2].

*Amazon shipment*

After an order has been placed with e-commerce giant Amazon, any subsequent wait time until delivery might be considered liminal in the negative sense of the word: there is little to be gained from the waiting and nothing to lose should the company be able to be more efficient. This situation lacks all narrative desire and stands naked as transactional logistics. But should we look closer at this as a first-principle problem, many of us will remember waiting for a present during childhood as being the *best* and most memorable part of the present.

We could create a Narrative Desire that mirrors that emotion. Likely we would first designate what type of orders we want to apply this to, and what orders we don't really desire but simply deem a necessary evil. For *desired products*, the metaphor of sexual foreplay - another liminal space - might be even more useful: the creation of desire through withholding. We should consider capturing this by revealing gradually more about the product and its attractive qualities and features as the package makes its way closer to delivery (no pun intended). It is not the reveal of new product features that is at the core of creating desire, but the playful hints of what has yet to arrive that rekindle desire when coupled directly to the wait and the tracked delivery process though the logistics system.

*Commodity retail*

With over 30,000 stores worldwide, we chose the recent commoditization of Starbucks' coffee as the targeted problem. Stores are designed for effective throughput and commerce, not meaningful experiences. We don't necessarily have to reverse the streamlined logistics of making, receiving, and paying for coffee. But we do have to create new expectations and liminality. We could, for example, *slow down time* to create enough sensory violation: semi-transparent windows holding the stores as liminal borders that also play out film scenes at half speed showing people from around the world calmly drinking coffee, watching each other and the world pass by. Perhaps reading. Abstracting the slow imagery as monochrome would further its role as ambience, not content there to distract or, like all else in the stores with bright and colorful screens, pander for attention. The difference between our approach and previous similar ones, for example, Slow Technology (Hallnäs and Redström, 2001) is our clear target in a specific Narrative Desire on the other side of liminality and not just the holding of attention.

---

[2] Eating in the Dark, SpotCoolStuff Travel, http://travel.spotcoolstuff.com/unusual-restaurants-eating-in-the-dark, accessed 14 June 2022

Secondly, with the core product of coffee itself we could play up the substance's history and qualities as a drug. The smell and darkness, its exotic origins, and the paraphernalia for making it are all easy to re-contextualize with branding, naming, packaging, and rituals borrowed from other illegal substances, especially those with long and colorful histories such as opium dens and ceremonies around ayahuasca. The targeted effect would be a heighten value by infusing the main product with mythological meaning and the drinking with a stronger sense of a desired narrative liminality: a treat for oneself, to break away from the noise and speed of the modern world; a small potential transformation through relaxed introspection and a quick phygital flirt with brewed darkness.

*Metaverse*
Considering the vast and profoundly undefined nature of commercial metaverses as application, it is potentially most interesting to explore how liminality can work in a space that is already so artificial and, arguably, liminal. That said, metaverses are not inherently transformational. We can play an immersive video game and be acutely present in that specific universe without feeling the need to change anything about ourselves.

The Liminal Design model can apply the same way as in our other examples, but with one expectation: the suspension of disbelief must consider what universe we are leaving behind - not just the one we left to enter the metaverse, but also the metaverse we come from at time of entry. Like a film's or book's story within a story, the separate liminal space inside the metaverse has all the powers to make us be fully present in a new narrative but also requires us to look at the specific border we are asking someone to traverse, not just a border in general. Our suggestions here echo J.G. Ballard's 1962 manifesto "Which Way to Inner Space?", advocating a shift from outer to inner space.

The transformational opportunities inside a metaverse, although unlikely to be a priority for commercial interests, are significant. VR and AR are already used for mental health interventions and have shown consistent results in clinical trials (Riva et al., 2016). It is unlikely, however, that deep interconnectedness between humans will be lastingly generated simply by meeting as avatars compared to the more involved example of remote presence given above. However, if a liminal place inside the metaverse offered a safe place to discuss and explore, for example, addiction, loss, or depression - that liminal space would offer a similar liminality as a support group for care workers or Alcoholics Anonymous meetings.

This, however, still doesn't take full advantage of the metaverse's unique digital potential: what we know of participants and their actions, the liminal space's ability to morph and to optimize the virtual world and alter interactions based on the global design target coupled with a fluid, local in-the-moment user-level story, and overarching narrative and real-time feedback. A perhaps radical, but technically possible and narratively interesting part of our Meta Liminality could explore the fluidity applied to the concept of other "actors" in the space. For example, we might mix and only partially disclose which participants are regular users and which are trained actors, experts, or digital agents. This creates another level of liminality that both breaks and holds what theater-arts refer to as the "fourth wall". This type of meta-theater inside a given narrative would be conducive to highly transformative experiences - the focus of this paper - because our sense of self is challenged on several levels at the same time.

In addition, the digital realism possible in a VR metaverse does offer other unique possibilities for generating empathy (Herrera et al., 2018). There is a key distinction between two types of *cognitive empathy* - the difference between "imagine-self" and "imagine-other" perspective taking. The latter, requiring more active participation, is akin to Marshall McLuhan's "cold media'" (McLuhan, 1964). We argue that cold media is likely to create deeper empathy here, for example by putting the user in "someone else's shoes". A graphic example to showcase the potency of this in the metaverse would be to see the cows' perspective through a raw, loud and intense "first person view" of what the final 10 minutes is like at a slaughterhouse. What this idea lacks in commercial prospect, it likely makes up for in its power to create very real empathy.

**Conclusion**

Liminal Design can provide an approach to work with design spaces that are characterized by inherent ambiguous and transformative qualities. While traditional structured approaches to experience design reduce complexity as a means to deal with the ephemeral nature of human experience, Liminal Design chooses another approach: In a structured way, it explores experiential design while embracing the impalpable, incorporeal, and transformative nature of deep real-life human experiences. There is a need for these kinds of approaches to support the practice of design for experiences that extend beyond those dictated by efficiency and simple pleasure. The Liminal Design model is a first attempt to address transformative experiences; it will require further work to mature. Nonetheless, it's essence, the design for undefined and contradictory in-between states that allow and stimulate us to consider new ways of being, addresses an urgent deficiency in the current landscape of commercial design.

The concepts used in this paper can make it sound like we are designing a new religion or a vending machine for Stendhal syndrome. But the design task in Liminal Design inevitably includes a level of transcendence. Big or small, the very same principles and dynamics hold true. The technology and the product we design might be very commercial, but our ultimate design task is always far from it and striving for the sublime.

This is also why Liminal Design is likely to be a struggle for many of the big tech companies favoring the lowest common denominator to brand access and technology platforms before suggesting any deeper experience that might be emotionally complex and highly personal. The fundamental requirement that any Narrative Desire must promise something truly meaningful to work further complicates simple engagement with most corporations.

Had we asked someone 20 years ago what they hoped computers and the Internet would bring humanity by the time of this article, a survey of today's technological landscape would be sure to disappoint. Re-approaching technology through the lens of Liminal Design pries open more doors for development and innovation and fundamentally challenges today's transactional and commercial nature through the questions it asks.

We trust that the beautiful complexity that comes with Liminal Design delivers not only experiences that we have lost, but also aspirations for what we have never seen. Like all changes in behavior, it is not about bending small parts of a narrative, but rather providing new ones that speak of hope more directly to our imagination. There is no one single way to apply Liminal Design. And in its wider acceptance as part of product development, we hope that the multitude of solutions it offers will also lead to the unlimited inclusiveness of just as many profound experiences.

## Acknowledgments

The authors are thankful to Lotte Peeters, who created all the visualizations included in the article. Authors also wish to thank Dr. Alice Chirico and Milica Petrovic for providing valuable comments and suggestions, which helped to improve the manuscript.